\documentclass[twocolumn,showpacs,preprintnumbers,amsmath,amssymb]{revtex4}

\usepackage{graphicx}
\usepackage{dcolumn}
\usepackage{bm}
\usepackage{epsfig}

\begin{document}


\title{Conditional preparation of single photons for scalable quantum-optical networking}

\author{Alfred B. U'Ren}
\author{Christine Silberhorn}
\author{Konrad Banaszek}
\author{Ian A. Walmsley}
\affiliation{Clarendon Laboratory, Oxford University, Parks
Road, Oxford, OX1 3PU, England.}

\date{\today}

%
\newcommand{\epsfg}[2]{\centerline{\scalebox{#2}{\epsfbox{#1}}}}

\begin{abstract}
A fundamental prerequisite for the implementation of linear
optical quantum computation is a source of single-photon
wavepackets capable of high-visibility interference in scalable
networks. These conditions can be met with micro-structured
waveguides in conjunction with ultra-short classical timing
pulses. By exploiting a novel type-II phasematching configuration
we demonstrate a waveguided
 single photon source exhibiting a conditional \textit{detection} efficiency exceeding
$51\%$ (which corresponds to a \textit{preparation} efficiency  of
$85\%$) and extraordinarily high detection rates of up to
$8.5\times10^5\mbox{coincidences/[s$\cdot$mW]}$.
\end{abstract}

\maketitle


Single photons provide an important bridge between classical and 
non-classical physics. For example, it is possible to define a 
single-photon wavefunction that has exactly the same form as the 
classical electromagnetic field\cite{bialynincki96}, yet the 
quadrature phase-space representations of the state can be singular 
or non-positive\cite{lvovsky01}. Apart from fundamental interest, the
ability to generate single-photon
wavepackets in a scalable manner is a prerequisite for the further
development of quantum-enhanced technologies. Recent progress in 
quantum information processing highlights the
necessity for a reliable single photon source. At the heart of
such novel proposals is quantum interference,
which necessitates photonic wavepackets exhibiting well-defined
photon number and a well-defined modal character. Thus, modal
distinguishability hinders the implementation of linear optical
quantum computation \cite{knill01,ralph01,pittman03} and indeed of
all schemes relying on interference between single photons from
multiple sources\cite{uren03} such as
teleportation\cite{bouwmeester97,boschi98}, entanglement
swapping\cite{pan98}  and networking via quantum
repeaters\cite{briegel98}. Furthermore, single photon emission in
well-defined modes permits efficient fiber coupling, crucial for
long-haul quantum cryptography and communication\cite{gisin02}. 

Two distinct approaches for generating single photons are
currently being pursued: deterministic sources of single photons
emitted on-demand, and spontaneous sources based on photon-pair
generation where a single photon is prepared by detection of the
conjugate pair member.  Sources based on single vacancy
centers\cite{kurtsiefer00}, quantum
dots\cite{michler00,yuan02,santori02}, atoms in
cavities\cite{kuhn02} and molecular emission\cite{lounis00} emit
photons deterministically, and often rely on intricate
experimental setups (e.g. cryostatic cooling). For solid-state
sources, however, it remains challenging to control the emission
modes, resulting in poor interference, poor fiber-coupling and low
detection efficiencies. This in turn leads to a random selection
of collected photons. In the process of parametric downconversion
(PDC), on the other hand,  photon pair emission occurs randomly
but the presence of  a single photon can be determined by the
detection of its sibling.  It is nevertheless difficult to collect
the entire photon sample from bulk crystals\cite{barberi03,yamamoto03} due to
the relatively complicated spatial emission pattern.  PDC from quasi
phasematched non-linear waveguides has recently been shown,
however, to exhibit emission in controlled modes defined by the
guide\cite{banaszek01,tanzilli01,booth02,sanaka01}. Accurate
spatial mode definition leads to efficient optical fiber coupling
and to much improved conditional detection rates as well as
high-visibility interference.  A fundamental requirement for
high-fidelity conditional preparation of single photons based on
waveguided PDC is efficient pair-splitting, which is realized here
through a nonlinear interaction producing orthogonally-polarized
(and therefore spatially separable) photon pairs.

The difficulty in generating orthogonally polarized PDC light in a
waveguided $\chi^{(2)}$ interaction is that existing waveguide
structures are commonly designed to take advantage of the high d$_{33}$ nonlinearities of LiNbO$_3$ and KTiOPO$_4$ (KTP) which implies the use of type-I phasematching
yielding same-polarization PDC photon pairs. Since waveguided PDC
additionally implies that both PDC photons in a given pair occupy a waveguide-supported
spatial mode, it is challenging to split the pairs.   For common
$\chi^{(2)}$ materials such as periodically poled
LiNbO$_3$ (PPLN), waveguiding supports only one polarization. Thus, to date, quantum optical experiments making use of nonlinear
waveguides have employed this type of
phasematching\cite{banaszek01,tanzilli01,booth02,sanaka01,anderson95}.
We have designed a type-II PDC interaction in a
periodically poled KTP waveguide
which leads to easily separable (by means of their polarization)
photon pairs. In such a phasematching configuration (utilizing the
$\chi^{(2)}$ element $d_{24}$), a horizontally-polarized
ultraviolet photon spontaneously decays into two infrared photons,
horizontally and vertically polarized.

\begin{figure}[ht]
\epsfg{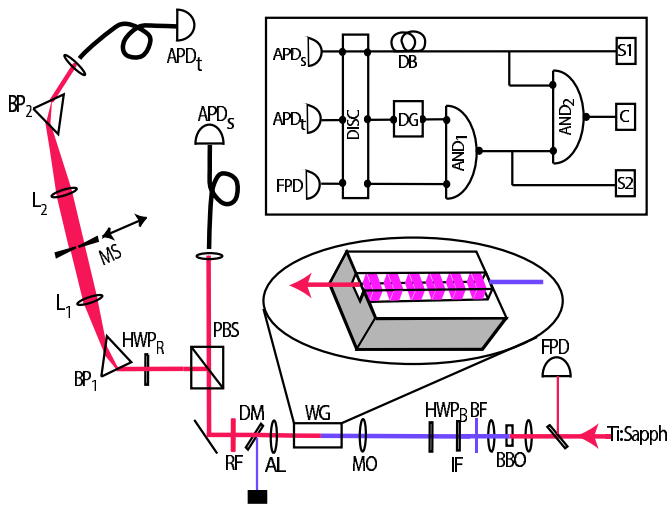}{.85} \caption{Experimental apparatus with photon-counting electronics,
including logic elements used for time-gating shown in inset.  A
quasi-phasematched nonlinear waveguide is set up to produce
orthogonally-polarized photon pairs via type-II downconversion.  A
polarization beam splitter divides the photon pair sample into two
modes (trigger and signal) one of which (trigger) is subjected to
spectral filtering and to post-detection time-gating (see inset)
while the signal mode is directly detected.  FPD: fast photodiode;
BBO: $2\mbox{mm}$ $\beta$-barium-borate doubling crystal; BF:
BG-39 Schott colored filter, IF: narrow-band pass filter, HWP$_B$:
half waveplate set to flip polarization; MO: 10X microscope
objective; WG: 12mm long KTiOPO$_4$ waveguide with $8.7\mu m$
period; AS: AR coated f=8mm aspheric lens; DM: blue-reflecting,
red-transmitting dichoric mirror; RF: AR-coated RG-665 Schott
colored filter; HWP$_R$ AR-coated halfwave plate; PBS: polarizing
beam splitter;  BP$_1$ and BP$_2$: Brewster-angle SF-10prism;
L$_1$ and L$_2$: $f=10\mbox{cm}$ AR-coated lens; MS: translatable
slit; APD$_t$: trigger fiber-coupled avalanche photodiode (APD)
from Perkin-Elmer; APD$_s$: signal APD; INV: pulse inverter; DISC:
pulse discriminator; DB: electronic variable delay line; DG:
electronic delay generator (Stanford research DG-535); AND$_1$ and
AND$_2$: NIM AND gates; S$_1$, S$_2$ and C: pulse counters.} \label{Fi:stackedhomis}
\end{figure}

\begin{figure}[ht!]
\centering
\includegraphics{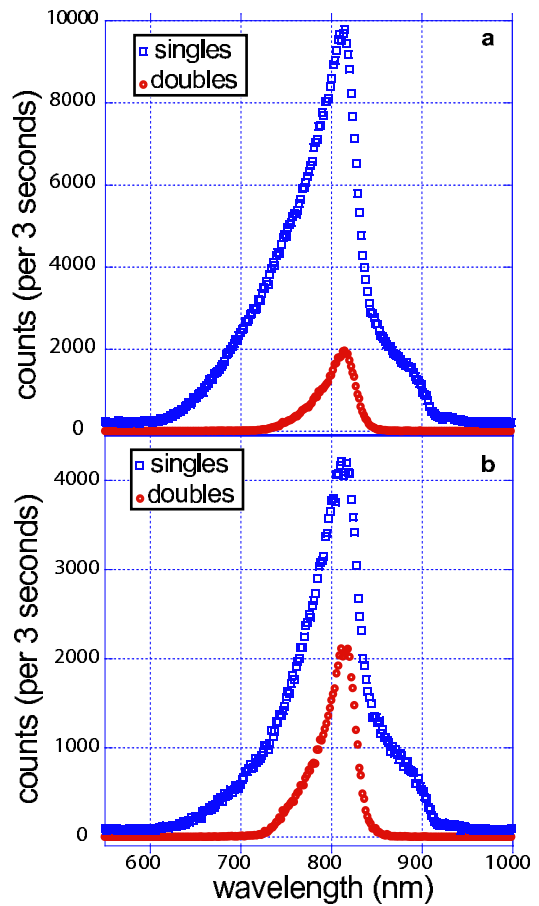}
\caption{Spectrally resolved parametric downconversion. (a)
Frequency-resolved singles (trigger) and coincidence counts without
time-gating. (b) Frequency-resolved singles (trigger) and coincidence counts with
time-gating.  Note that the maximum conditional detection
efficiency increases from $20.4\%$ to $50.5\%$ upon activation
of time-gating. }
\end{figure}

Our experimental apparatus is depicted in Fig. 1.  The output of a
mode-locked titanium sapphire laser ($100\mbox{fs}$ pulse
duration, 87MHz repetition rate) is directed to a 2mm-long
$\beta$-barium-borate crystal yielding pulses centered at
$400.5\mbox{nm}$, whose bandwidth is restricted by an interference
filter with a FWHM of $2\mbox{nm}$. This ultraviolet beam (power
measured before coupling was ~$15 \mu \mbox{W}$) is focused using
a 10X microscope objective into the input face of a
$12\mbox{mm}$-long periodically poled z-cut KTP-waveguide (with
$8.7\mu \mbox{m}$ grating period). The waveguide output is
collimated and the remaining ultraviolet is filtered out from the
PDC signal. The photon pairs are subsequently split by a
polarizing beam splitter and the horizontally polarized signal
mode is coupled by a multi-mode fiber to a commercial
silicon-based avalanche photodiode (APD). The trigger channel
(vertical polarization) is subjected to a low-loss prism
spectrometer comprised of two SF-10 Brewster angle prisms and two
$f=10\mbox{cm}$ lenses; a motorized slit of adjustable width is
placed at the Fourier plane whose position is computer-controlled.
The resulting trigger mode is similarly launched into a
fiber-coupled APD.  The slit position is calibrated by
transmitting a titanium sapphire beam through the prism setup into
a spectrometer, where a linear extrapolation was made for
wavelengths outside the laser bandwidth.  To implement
time-gating, a small percentage of the laser power is directed to
a fast photodiode (1ns rise-time).  The diode signal is amplified
and discriminated producing a train of $3\mbox{ns}$ duration
pulses which is delayed (with an electronic delay generator) and
combined at an AND gate with the discriminated trigger output.  A
second AND gate comparing the time-gated and non-gated signals
registers when both inputs arrive simultaneously (to within
$3\mbox{ns}$).

Waveguided PDC leads to several key advantages: it increases
source brightness and enables conditional preparation efficiencies
limited only by detector losses while attaining accurate
spatio-temporal modal control. Despite the fact that in KTP the
$d_{24}$ element is considerably smaller in magnitude than those
elements yielding type I PDC ($d_{33}$ and $d_{31}$), it is
possible to obtain a remarkably high production rate of type-II
PDC photon pairs since the nonlinear gain exhibits a quadratic
dependence on the guide length (rather than linear dependence for
bulk crystals).  In our experiment one of the polarizations is
regarded as a trigger, while we attempt to collect all photons in
the orthogonal polarization (signal). In the limit of unit
detector quantum efficiency together with vanishing optical losses
and perfectly-suppressed background, a trigger detection heralds
the presence of a single photon in the signal arm.   Such
conditional detection is characterized by an efficiency given by
the ratio of coincidence (trigger and signal) to singles (trigger)
detection rates. The source brightness, given by the coincidence
rate per unit pump power, is an additional important measure of
source performance, specifically in the context of concatenating
multiple waveguides for quantum-optical networking.

We encountered two important sources of background photons
produced by our waveguide. If not suppressed, such uncorrelated
light is a serious limitation to the conditional detection
efficiency. First, the quasi phasematched grating needed for
type-II PDC (with $8-10 \mu m$ period) also supports type-I PDC,
resulting from the $\chi^{(2)}$ elements $d_{33}$ and $d_{31}$,
producing same-polarization pairs which do not contribute to
coincidence events and thus reduce the conditional detection
efficiency. Fortunately, the various phasematched processes in
the waveguide are spectrally distinct; indeed, we verified that a
band-pass filter in the path of the ultraviolet pump can suppress
type-I interactions.  Second, the waveguide produces uncorrelated
fluorescence photons, with an intensity comparable to that of PDC.
The observed fluorescence is related to gray-tracking in
KTP due to color-center
formation\cite{boulanger99} and has been observed in PDC from bulk
periodically poled material\cite{kuklewicz03}.   While in a
waveguide a substantial fraction of the fluorescence is emitted
into the supported modes, we found that the fluorescence and PDC
signals exhibit certain features that can be exploited to
differentiate between them. By direct measurement we determined
that the fluorescence spectrum is considerably wider than that of
PDC (130nm versus 50nm 1/e full width). By filtering out
frequencies at which PDC is not present, fluorescence is
suppressed without appreciably reducing the PDC photon sample.
Moreover, while PDC events occur within the femtosecond pump pulse
window, fluorescence is emitted over much longer timescales.
Therefore, gating in time with respect to the pump pulse train
leads to further fluorescence suppression.

An experimental run consists of the recording of singles and
coincidence detection rates as a function of the slit position. A
slit width of $40\mu \mbox{m}$ maximizes the transmitted signal at
the highest spectral resolution (about $2\mbox{nm}$).   Data was
taken with and without time-gating, as shown in Fig. 2.   The
maximum coincidence to singles ratio (i.e. the conditional detection efficiency)
increases from $20.4\%$ to
$51.5\%$ upon activation of time-gating.  If the heralded single photons are prepared for a
subsequent experiment, rather than detected directly, the
preparation efficiency does not include the signal detection loss,
which together with imperfect fiber coupling is the highest source
of loss (detector specifications indicate $60\%$ quantum
efficiency at $800\mbox{nm}$, measured fiber coupling efficiencies
were $>90\%$ while all other optics incur negligible loss).  For a
quantum efficiency of $60\%$, a $51.5\%$ conditional detection efficiency corresponds
to a preparation efficiency
of single photons of close to $85\%$.  The latter means that we can ascertain the presence of
a single photon in a well-defined spatio-temporal mode with an
$85\%$ fidelity.  Furthermore, annealing the color centers,
e.g. by heating the waveguide\cite{boulanger99}, may suppress the
remaining fluorescence and lead to nearly ideal single-photon
preparation.

\begin{figure}[h!]
\centering
\includegraphics{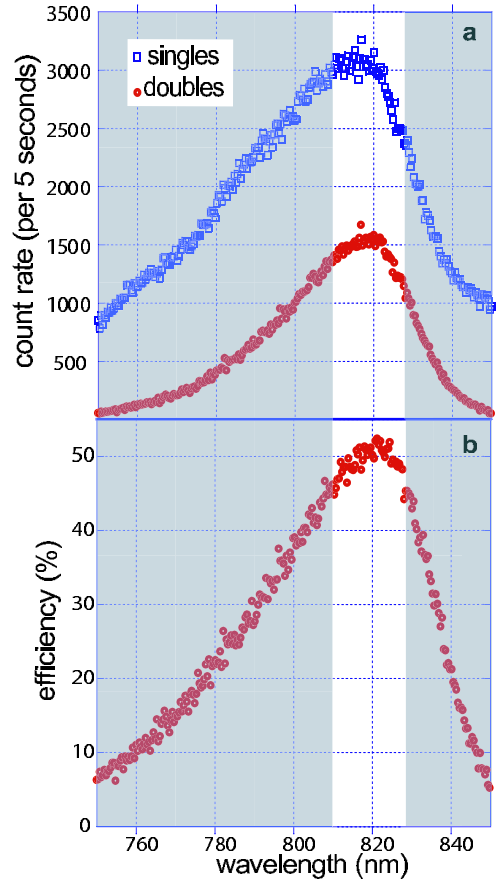}
\caption{Conditional detection efficiency for optimized source
brightness. This figure shows spectrally-resolved coincidences and
singles counts in the region of the coincidences peak. The
non-shaded band indicates the location and width (corresponding to
a $17\mbox{nm}$ spectral transmission window) of the pump
spectrometer slit yielding the highest brightness ($8.5 \times
10^5$ coincidences/[s$\cdot$ mW]) at the maximum conditional
detection efficiency ($~51\%$). (a) Depicts the
frequency-resolved coincidence and singles (trigger) detection rates (b)
Depicts the conditional detection efficiency (given by the ratio
of coincidence to singles counts).}
\end{figure}

In a second experiment, we optimized the source
brightness in order to maximize optical
throughput while retaining a high efficiency by adjusting the slit
position and width, resulting in
a $17\mbox{nm}$ transmission window. For a $300\mbox{s}$
integration time, we observed $7.46\times 10^5$ trigger,
$1.15\times 10^7$ signal and $3.81\times 10^5$ coincidence counts
(the calculated accidental coincidence rate is $<330$ counts) corresponding
to a brightness of $8.5\times 10^5
\mbox{coincidences/(s$\cdot$ mW)}$.  For comparison with our
spectrally-resolved measurements, Fig.~3 shows experimental data
close to the coincidence peak; the non-shaded band indicates the
slit position and width corresponding to simultaneous brightness
and efficiency maximization.  We have thus shown the experimental
realization of high-fidelity conditional preparation of
fiber-coupled single photons generated by a femtosecond-pulse
pumped waveguide (in a micro-structured nonlinear optical array)
based on orthogonally-polarized parametric downconversion photon pair
generation from a periodically poled KTiOPO$_4$ nonlinear
waveguide\cite{roelofs94}. The use of waveguiding leads to a high
probability of photon pair generation which translates for our experiment into an
extraordinarily high detection rate and a remarkable conditional
efficiency. Such
single photons are characterized by an ultrashort wavepacket with a broad
spectral bandwidth.
Furthermore, the use of an ultrashort pump pulse train constrains
emission times to within a femtosecond-duration window, crucial
for applications requiring synchronized emission from multiple
sources.

To our knowledge, the best previously reported ratio obtained
with CW-pumped PDC from a $\beta$-barium-borate crystal and
collected with single-mode fibers was $28.6\%$ at a brightness of
$775\mbox{ counts/(s$\cdot$ mW)}$\cite{kurtsiefer01}. The use of a CW pump in this experiment implies that there is not a classical timing signal for synchronization of multiple sources.  Experiments
aimed at determining the quantum efficiency of single-photon
detectors have reported high coincidence to singles
ratios\cite{kwiat93} when corrected for optical losses; however,
photons in the signal arm lacked modal definition, limiting the
potential for usable conditionally-prepared single photons. We
believe that our higher brightness arises from accurate modal
definition at the source, leading to efficient fiber-coupling of
the whole photon sample. In contrast, for bulk crystal PDC, mode
definition is only possible \textit{a posteriori} (e.g. with irises or
fibers).

Further development of our source could include modal engineering
of the conditionally-prepared photon states to yield well-defined
pure photon-number states (i.e. Fock states) and arbitrary
superposition wavepackets\cite{grice01,dakna99,pegg98}.
Mode-matching into single-mode fibers should be straightforward
given that the waveguide exhibits accurate spatial mode control.
Precise timing together with high brightness paves the road
towards concatenation of multiple waveguides in integrated
quantum-optical networks. Thus our observed high brightness
together with the use of an ultra-short pump, leads to source
scalability by utilizing multi-waveguide chips. In addition,
utilizing a higher pump power such high brightness permits the
generation of higher-occupancy Fock states at
experimentally-usable production rates.   In conclusion, our
source is an ideal building block for quantum information
applications offering compatibility with all-fiber systems, while
room-temperature operation makes it a convenient alternative to
solid-state sources.


\end{document}